\documentstyle[12pt]{article}
\def\be{\begin{equation}}
\def\ee{\end{equation}}
\def\bea{\begin{eqnarray}}
\def\eea{\end{eqnarray}}

\input{tcilatex}
\begin{document}

\begin{titlepage}

\title{Multisoliton solution of 3-waves problem}

\author{A.N. Leznov$^{1}$, G. R. Toker$^{1}$ and R.Torres-Cordoba$^{2}$\\  $^1${Universidad Autonoma del Estado de Morelos, CIICAp}\\ 
$^2${Universidad Autonoma de Ciudad Juarez, ITT}}

\maketitle
\begin{abstract}

Multi-soliton solution of the 3-waves problem is represented in explicit determinative form. 

\end{abstract}

\end{titlepage}

\section{Introduction}

The problem of three waves in two dimensions arises in different forms in
many branches of the mathematical physics. For example, this occurs in radio
physics and nonlinear optics applications, which can be found in \cite{1}.

Three-wave interaction in plasma, stability criteria, and asymptotic
behavior for a general system of three interacting waves, the influence of
mutually different linear damping coefficients on a system of three
interacting waves of equal signs of energy are described in \cite{A}, where
an accurate studying waves in plasma were performed.

Nonlinear Water Wave Interaction and hydrodynamic turbulence including tree
wave problem were investigated in \cite{B, C, D}

Furthermore, the three wave problem is also discussed during studying
nonlinear lattices \cite{F} and photon crystals \cite{G}.

The present paper must be considered as a direct continuation of the
previous one \cite{I}. For the convenience of the reader we at first
represent necessary results from \cite{I} marking them into the form of an
Appendix, putting this on the unequal first place.

\section{Appendix}

\subsection{3-wave problem in dimensionless form and its discrete
transformations.}

First off, let us consider the system of equations for 6 unknown functions $%
Q,P,A,D.B.E$ 
\[
P_{1}=-QE,\quad A_{2}=-BQ,\quad -(Q_{1}+Q_{2})\equiv Q_{3}=-PA 
\]
\begin{equation}
B_{1}=-AD,\quad E_{2}=-DP,\quad D_{3}=-EB  \label{MS}
\end{equation}

Using direct calculation it is not difficult to certain that system (\ref{MS}
is invariant with respect to the following substitution: $T_{3}$) 
\[
\bar{Q}={\frac{1}{D}},\quad \bar{A}=-{\frac{B}{D}},\quad \bar{P}={\frac{E}{D}%
}, 
\]
\[
\bar{B}=D({\frac{B}{D}})_{2},\quad \bar{E}=-D({\frac{E}{D}})_{1},\quad {%
\frac{\bar{D}}{D}}=DQ-(\ln D)_{1,2} 
\]
The transformation is denoted as $T_{3}$, which is a discrete transformation.

Permutating indexes $(1,3)$ (altogether with corresponding exchanging
unknown functions) make it possible to obtain $T_{1}$ as a discrete
transformation when the system (\ref{MS}) is invariant. 
\[
\bar{P}={\frac{1}{B}},\quad \bar{Q}={\frac{A}{B}},\quad \bar{E}=-{\frac{D}{B}%
}, 
\]
\[
\bar{D}=B({\frac{D}{B}})_{2},\quad \bar{A}=-B({\frac{A}{B}})_{3},\quad {%
\frac{\bar{B}}{B}}=BP-(\ln B)_{2,3} 
\]
And finally the discrete transformation $T_{2}$ has the form of: 
\[
\bar{A}={\frac{1}{E}},\quad \bar{B}={\frac{D}{E}},\quad \bar{Q}=-{\frac{P}{E}%
}, 
\]
\[
\bar{D}=-E({\frac{D}{E}})_{1},\quad \bar{P}=E({\frac{P}{E}})_{3},\quad {%
\frac{\bar{E}}{E}}=EA-(\ln E)_{1,3} 
\]

In the form presented above, substitutions $T_{i}$ may be considered as a
mapping connected to six initial (unbar) functions with six final (bar)
ones. On the other hand, each substitution should be considered as the
infinite dimensional chain of equations. For instance, the corresponding
chain of equations for the case of $T_{1}$ substitution has the form of: 
\begin{equation}
{\frac{B^{n+1}}{B^{n}}}-{\frac{B^{n}}{B^{n-1}}}=-(\ln B^{n})_{2,3},\quad
D^{n+1}=B^{n}({\frac{D^{n}}{B^{n}}})_{2},\quad A^{n+1}=-B^{n}({\frac{A^{n}}{%
B^{n}}})_{3}  \label{CH}
\end{equation}
\[
E^{n+1}=-{\frac{D^{n}}{B^{n}}},\quad Q^{n+1}={\frac{A^{n}}{B^{n}}} 
\]
In the first row, we have the lattice as if it were system connected to 3
unknown functions $(B,D,A)$ in each point of the lattice. The first chain
for $B$ functions is definitely well known as a two dimensional Toda lattice.

\subsection{Some properties of the discrete transformations.}

All of the discrete transformations constructed above are invertable. This
means that an unbar unknown function may be presented in terms of the bar
ones. For instance, $T_{3}^{-1}$ looks as: 
\[
D={\frac{1}{\bar{Q}}},\quad B=-{\frac{\bar{A}}{\bar{Q}}},\quad E={\frac{\bar{%
P}}{\bar{Q}}}, 
\]
\[
P=-\bar{Q}({\frac{\bar{P}}{\bar{Q}}})_{2},\quad A=\bar{Q}({\frac{\bar{A}}{%
\bar{Q}}})_{1},\quad {\frac{Q}{\bar{Q}}}=\bar{D}\bar{Q}-(\ln \bar{Q})_{1,2} 
\]

Consequently, it is not difficult to certain by direct computation that
discrete transformations $T_{i}$ are mutual commutative $%
(T_{i}T_{j}=T_{j}T_{i})$ in the solutions of the system (\ref{MS}).
Moreover, others, such as $T_{1}T_{2}=T_{2}T_{1}=t_{3}$, can be also handled
in the same way.

Thus, from each given initial solution $W_{0}\equiv (A,P,Q,E,B,D)$ of the
system (\ref{MS}), it is possible to obtain the chain of solutions labeled
by two natural numbers $(l_{1},l_{2}$, or $(l_{3}))$ and the number of
applications of the discrete transformations $(T_{1},T_{2},T_{3})$
associated to it (as it was noticed above, $T_{1}T_{2}=T_{2}T_{1}=T_{3}$).

Arising chain of equations with respect to $(D,B,E)$ functions are
definetely two-dimensional Toda lattices. Their general solutions in the
case of two fixed ends are well-known \cite{LS}. As the reader will see,
soon this fact allows constructing the many soliton solutions of the 3-wave
problem into the most straightforward ways.

\subsection{Resolving of discrete transformation chains.}

\subsubsection{Two identities of Yacobi}

The first Jacobi identity: 
\[
D_{n}\pmatrix{T_{n-1}a^{1}\cr
b^{1}\tau _{11}\cr}D_{n}\pmatrix{T_{n-1}a^{2}\cr
b^{2}\tau _{22}\cr}-D_{n}\pmatrix{T_{n-1}a^{2}\cr
b^{1}\tau _{12}\cr}D_{n}\pmatrix{T_{n-1}a^{1}\cr
b^{2}\tau _{21}\cr}= 
\]
\[
D_{n-1}(T_{n-1})D_{n+1}\pmatrix{T_{n-1}a^{1}a^{2}\cr
b^{1}\tau _{11}\tau _{12}\cr
b^{2}\tau _{21}\tau _{22}\cr} 
\]
where $a^{i},b^{i}$ are $(n-2)$ dimensional columns (rows) vectors, $\tau
_{i,j}$ components of 2-th dimensional matrix.

The second Jacobi identity: 
\[
D_{n}\pmatrix{T_{n-1}a^{1}\cr
b^{1}\tau \cr}D_{n+1}\pmatrix{T_{n-1}a^{1}a^{2}\cr
d^{1}\nu \mu \cr
b^{2}\rho \tau \cr}-D_{n}\pmatrix{T_{n-1}a^{1}\cr
b^{2}\rho \cr}D_{n+1}\pmatrix{T_{n-1}a^{1}a^{2}\cr
d^{1}\nu \mu \cr
b^{1}\tau \sigma \cr}= 
\]
\[
D_{n}\pmatrix{T_{n-1}a^{1}\cr
d^{1}\nu \cr}D_{n+1}\pmatrix{T_{n-1}a^{1}a^{2}\cr
b^{2}\rho \tau \cr
b^{1}\tau \sigma \cr} 
\]

These identities can be generalized in the case of arbitrary semi-simple
group. The reader can find these results in \cite{3}.

\subsubsection{Resolving of the discrete lattices}

Let us take an initial solution in the form of: 
\begin{equation}
Q=A=P=0,\quad B\equiv B(2),\quad E\equiv E(1).\quad D_{3}=-BE  \label{IS}
\end{equation}
Application to this solution to each of the inverse transformations $%
T_{i}^{-1}$ is meaningless because of arising zeroes in denominators. The
chain of equations under such boundary conditions is what we call the chain
with the fixed end from the left (from one side).

The result of applications to such initial solution by $l_{3}$ times $T_{3}$
transformation looks as (in order for checking this fact only two Yacobi
identities of the previous subsection are necessary). 
\[
Q^{(l_{3}}=(-1)^{l_{3}-1}{\frac{\Delta _{l_{3}-1}}{\Delta _{l_{3}}}},\quad
D^{(l_{3}}=(-1)^{l_{3}}{\frac{\Delta _{l_{3}+1}}{\Delta _{l_{3}}}},\quad
\Delta _{0}=1 
\]
\begin{equation}
A^{(l_{3}}=(-1)^{l_{3}}{\frac{\Delta _{l_{3}}^{B}}{\Delta _{l_{3}}}},\quad
P^{(l_{3}}={\frac{\Delta _{l_{3}}^{E}}{\Delta _{l_{3}}}},\quad \Delta
_{0}^{B}=\Delta _{0}^{E}=0  \label{T_3}
\end{equation}
\[
B^{(l_{3}}={\frac{\Delta _{l_{3}+1}^{B}}{\Delta _{l_{3}}}},\quad
E^{(l_{3}}=(-1)^{l_{3}}{\frac{\Delta _{l_{3}+1}^{E}}{\Delta _{l_{3}}}},\quad
\Delta _{-1}=0. 
\]
where $\Delta _{n}$ are minors of the n-th order of infinite dimensional
matrix 
\begin{equation}
\Delta =\pmatrix{ D & D_2 & D_{22} & .....\cr D_1 & D_{12} & D_{122} &
.....\cr D_{11} & D_{112} & D_{1122} & .....\cr ...... & ....... &
.........& .....\cr}  \label{DM}
\end{equation}
and $\Delta _{l_{3}}^{E},\Delta _{l_{3}}^{B}$ are the minors of $l_{3}$
order in the matrices of which the last column (or row) is exchanged on the
derivatives of the corresponding order on argument $1$ of $E$ function (on
argument $2$ of the $B$ function in the second case).

In the following next notations will be used. $W^{l_3,l_1}$, ($W^{l_3,l_2}$%
)- this is the result of application of discrete transformation $%
T^{l_3}T^{l_1}$ ($T^{l_3}T^{l_2}$) to the corresponding component of the
3-wave field. $\Delta_{l_3,l_1}$ ($\Delta_{l_3,l_2}$) - determinant of $%
l_3+l_1$ ($l_3+l_2$) orders, with the following structure of its determinant
matrix. The first $l_3$ rows (columns) of it coinside with matrix of (\ref
{DM}) and last $l_1$, $(l_2)$ rows (columns)are consructed from the
derivatives of $B$, ($E$) functions with respect to arguments 2, (1).

The result of additional application to $l_{1}$ times $T_{1}$ transformation
to the solution (\ref{T_3}) looks as : 
\[
P^{l_{3},l_{1}}={\frac{\Delta _{l_{3},l_{1}-1}}{\Delta _{l_{3},l_{1}}}}%
,\quad B^{l_{3},l_{1}}={\frac{\Delta _{l_{3},l_{1}+1}}{\Delta _{l_{3},l_{1}}}%
},\quad \Delta _{0}=1,\quad \Delta ^{l_{3},-1}\equiv \Delta _{l_{3}}^{E} 
\]
\begin{equation}
Q^{l_{3},l_{1}}=(-1)^{l_{3}+l_{1}-1}{\frac{\Delta _{l_{3}-1,l_{1}}}{\Delta
_{l_{3},l_{1}}}},\quad D^{l_{3},l_{1}}=(-1)^{l_{3}+l_{1}}{\frac{\Delta
_{l_{3}+1,l_{1}}}{\Delta _{l_{3},l_{1}}}},  \label{TT}
\end{equation}
\[
E^{l_{3},l_{1}}=(-1)^{l_{3}+l_{1}}{\frac{\Delta _{l_{3}+1,l_{1}-1}}{\Delta
_{l_{3},l_{1}}}},\quad A^{l_{3},l_{1}}=(-1)^{l_{3}+l_{1}}{\frac{\Delta
_{l_{3}-1,l_{1}+1}}{\Delta _{l_{3},l_{1}}}}, 
\]
We do not present the explicit form for components $W^{(}{l_{3},l_{2}}$,
which can be obtained without any difficulties from (\ref{TT}) by
corresponding exchaging of the arguments and unknown functions.

\section{Multi-soliton solution of the 3-waves problem}

\subsection{General consideration}

The system (\ref{MS}) allows reducing (under additional asumption that all
operators of differentiation are the real ones $\partial _{\alpha }=\partial
_{\alpha }^{*}$) . 
\begin{equation}
P=B^{*},\quad A=E^{*},\quad Q=D^{*}  \label{REA}
\end{equation}
In this case the system (\ref{MS}) is reduced to three equations. 
\begin{equation}
B_{1}=-DE^{*},\quad E_{2}=-DB^{*},\quad D_{3}=-BE  \label{MSC}
\end{equation}
for three complex unknown functions $(E,B,D)$. This system of equations is
definetely 3-wave problem.

In \cite{I} it was shown that begining with an initial solution (\ref{IS})
after the $(l_{3},l_{1}(l_{2}))$ steps of discrete transformation leads to a
solution of the reduced system (\ref{MSC}) if the following conditions are
satisfied: 
\[
\Delta _{2l_{3}+1,2l_{1}}=\Delta _{2l_{3}+1,2l_{1}-1}=\Delta
_{2l_{3},2l_{1}+1}=0 
\]
\begin{equation}
{\frac{\Delta _{2l_{3}-1,2l_{1}+1}}{\Delta _{2l_{3},2l_{1}}}}%
=(-1)^{l_{3}+l_{1}}E^{*}\quad {\frac{\Delta _{2l_{3}-1,2l_{1}}}{\Delta
_{2l_{3},2l_{1}}}}=(-1)^{l_{1}-1}D^{*}\quad {\frac{\Delta _{2l_{3},2l_{1}-1}%
}{\Delta _{2l_{3},2l_{1}}}}=(-1)^{l_{3}}B^{*}  \label{AC}
\end{equation}
Resolving the first line of above equations is as follows: 
\[
D=\sum_{k=1}^{2l_{3}}f^{k}(1)\phi ^{k}(2), 
\]
To resolve the second line ones the following parametrization is sufficient: 
\[
\quad f^{k}=\sum_{s=1}^{2l_{3}}c_{s}^{k}e^{\lambda _{s}1},\quad \phi
^{k}=\sum_{S=1}^{2l_{3}+2l_{1}}d_{S}^{k}e^{\mu _{S}2}, 
\]
\[
E=\sum_{s=1}^{2l_{3}}e_{s}e^{\lambda _{s}1},\quad
B=\sum_{S=1}^{2l_{3}+2l_{1}}b_{S}e^{\mu _{S}2} 
\]
All numerical parameters $c_{s}^{k},d_{S}^{k},e_{s},b_{S}$ are connected by
equations (\ref{AC}) and one additional equation connects with initial
conditions (\ref{IS}).

The initial condition leads to: 
\[
(\lambda _{s}+\mu _{S})\sum_{k=1}^{2l_{3}}c_{s}^{k}d_{S}^{k}=e_{s}b_{S}\quad
d_{S}^{k}=b_{S}\sum_{s=1}^{2l_{3}}e_{s}{\frac{(c^{-1})_{s}^{k}}{\lambda
_{s}+\mu _{S}}} 
\]
and finally we obtain for $D:$ 
\begin{equation}
D=\sum_{s=1,S=1}^{2l_{3},2(l_{3}+l_{1})}{\frac{e_{s}e^{\lambda
_{s}1}b_{S}e^{\mu _{S}2}}{\lambda _{s}+\mu _{S}}}  \label{D}
\end{equation}
Thus, solution to be determined are defined by the pair of $(2l_{3})$
parmeters $e_{s},\lambda _{s}$ and $2(l_{3}+l_{1})$ pairs parameters $%
b_{S},\mu _{S}$. Equations (\ref{AC}) give some additional limitations (of
reality) for these parameters. Below, we find these limitations and then $%
l_{3},l_{1}$ is given by this explicit formula (\ref{TT}), in which
conditions of reducing are satisfied.

\subsection{Determinant camputation. The case of denumerator.}

At first, let us calculate the common determinant (see (\ref{AC})) -$\Delta
_{2l_{3},2l_{1}}$ using explicit expressions for initial functions $D,E,B$.
The determinant matrix of dimension $2(l_{3}+l_{1})\times 2(l_{3}+l_{1})$
may be presented as product of two matrices. The first one has the block
form $2l_{3}\times 2l_{3}$ matrix $L(e,\lambda ;1)$ in upper left angle and
zeroes on other places except of the unity on main diagonal. The elements of
the matrix $L$ are the following $L_{ik}=\lambda _{k}^{i-1}e_{k}e^{\lambda
_{k}1}$. The result is obvious: 
\begin{equation}
\Delta
_{2l_{3},2l_{1}}^{(1}=Det(L)=\prod_{s=1}^{2l_{3}}e_{s}W_{2l_{3}}(\lambda
_{1},...\lambda _{2l_{3}})e^{(\sum_{k=1}^{2l_{3}}\lambda _{k})1}  \label{TR}
\end{equation}
where $W$ is a Vandermond determinant. The second matrix in its turn can be
represented as product of another two ones. The first $2l_{3}$ lines of the
first matrix have the matrix elements ${\frac{1}{\lambda _{s}+\mu _{S}}}$ ($%
s $ number of line, $S$ number of column). The remaining $2l_{1}$ lines of
this matrix are the usual Vandermond matrices $W_{i,S}=\mu _{S}^{i-1}$.
Computation of this determinant is not complicate problem with the following
result: 
\[
\Delta _{2l_{3},2l_{1}}^{(2}=W_{2l_{3}}(\lambda _{1},...\lambda
_{2l_{3}})W_{2(l_{3}+l_{1})}(\mu _{1},...\mu
_{2(l_{3}+l_{1})})\prod_{s=1,S=1}^{s=2l_{3},S=2(l_{3}+l_{1})}{\frac{1}{%
\lambda _{s}+\mu _{S}}} 
\]
Finally, the matrix elements of last $2(l_{3}+l_{1})\times 2(l_{3}+l_{1})$
matrix have the following analitical structure $b_{i}e^{\mu _{i}}\mu
_{i}^{k-1}$. And in consequence: 
\[
\Delta
_{2l_{3},2l_{1}}^{(3}=\prod_{S=1}^{2(l_{3}+l_{1})}b_{S}W_{2(l_{3}+l_{1})}(%
\mu _{1},...\mu _{2(l_{3}+l_{1})})e^{(\sum_{S=1}^{2(l_{3}+l_{1})}\mu _{S})2} 
\]
Thus, for $\Delta _{2l_{3},2l_{1}}$ summating all results above we obtain: 
\begin{equation}
W_{2l_{3}}^{2}(\lambda _{1},...\lambda _{2l_{3}})W_{2(l_{3}+l_{1})}^{2}(\mu
_{1},...\mu _{2(l_{3}+l_{1})})\prod_{s=1,S=1}^{s=2l_{3},S=2(l_{3}+l_{1})}{%
\frac{1}{\lambda _{s}+\mu _{S}}}\prod_{s=1}^{2l_{3}}e_{s}e^{\lambda
_{s}1}\prod_{S=1}^{2(l_{3}+l_{1})}b_{S}e^{\mu _{S}2}  \label{YX}
\end{equation}

\subsection{Determinant computation. The case of $B$ function}

In the process of computation of $\Delta _{2l_{3},2l_{1}-1}$, the
determinant $(2(l_{3}+l_{1})-1)\times (2(l_{3}+l_{1})-1)$ matrix may be
represented as product of two matrices. The first one coincides with matrix $%
L$ of the previous subsection. Determinant of this matrix has been already
calculated in (\ref{TR}). Second matrix in its turn may be represented in
the form of the product of two rectangular matrices of dimentions $%
(2(l_{3}+l_{1})-1)\times 2(l_{3}+l_{1}))$ and $2(l_{3}+l_{1})\times
(2(l_{3}+l_{1})-1)$, correspondently. The structure of these matrices are
described in the previous subsection. The determinant of the degree $%
2(l_{3}+l_{1})-1$ is equal to sum of products of all minors of $%
2(l_{3}+l_{1})-1$ orders of both matrices. These determinants were computed
also in the previous subsection. The finally result is as follows: 
\[
\sum_{i=1}^{2(l_{3}+l_{1}}W_{2l_{3}}(\lambda _{1},...\lambda
_{2l_{3}})W_{2(l_{3}+l_{1})-1}^{2}(\mu _{1},..\mu _{i-1},\mu _{i+1},..\mu
_{2(l_{3}+l_{1})})\prod_{S=1,S\neq i}^{2(l_{3}+l_{1})}b_{S}e^{\mu
_{S}2}\prod_{s=1,S=1,\neq i}^{s=2l_{3},S=2(l_{3}+l_{1})}{\frac{1}{\lambda
_{s}+\mu _{S}}} 
\]
Taking into account the last equality from (\ref{AC}) and substituting in it
all the results for determinant calculations above, we obtain: 
\[
\sum_{S=1}^{2(l_{3}+l_{1})}\prod_{s=1}^{2l_{3}}(\lambda _{s}+\mu
_{S})\prod_{K=1,K\neq S}^{2(l_{3}+l_{1})}(\mu _{K}-\mu _{S})^{-2}e^{-\mu
_{S}2}{\frac{1}{b_{S}}}=(-1)^{-l_{3}}\sum_{S=1}^{2(l_{3}+l_{1})}b_{S}^{*}e^{%
\mu _{S}^{*}2} 
\]
The last equality can be satisfied only under condition: 
\[
\mu _{S}^{*}=-\mu _{PS},\quad S\to (PS) 
\]
where $P$ operator of permutation of $2(l_{3}+l_{1})$ numbers with obvious
propertity $P^{2}=1$. Comparing terms under the same exponents, we obtain: 
\[
(-1)^{l_{3}}b_{PS}^{*}=\prod_{s=1}^{2l_{3}}(\lambda _{s}+\mu
_{S})\prod_{K=1,K\neq S}^{2(l_{3}+l_{1})}(\mu _{K}-\mu _{S})^{-2}e^{-\mu
_{S}2}{\frac{1}{b_{S}}} 
\]
This is typical for soliton theory connection between the ''energy'' and
amplitudes.

\subsection{Determinant computation. The case of $E$ function.}

Numerical parameters $c_{s}$ are connected to the first equation (\ref{AC}).
To find these relations it is necessary to calculate $\Delta
_{2l_{3}-1,2l_{1}+1}$. The technique of these calculations are the same as
in previous subsection and finally equality (\ref{AC}) for $E$ functions
looks as: 
\[
\sum_{s=1}^{2l_{3}}e_{s}^{*}e^{\lambda _{s}^{*}1}=\sum_{s=1}^{2l_{3}}{\frac{1%
}{e_{s}}}e^{-\lambda _{s}1}\prod_{S=1}^{2(l_{3}+l_{1})}(\lambda _{s}+\mu
_{S})\prod_{k=1,k\neq s}^{2l_{3}}(\lambda _{k}-\lambda _{s})^{-2} 
\]
As in previous subsection, the equality above restricted values of
parameters $\lambda $ as: 
\[
\lambda _{s}^{*}=-\lambda _{Ps} 
\]
where $P$ {\bf is} some permutation of the group of $2l_{3}$ numbers
satisfying the condition $P^{2}=1$. Comparison the coefficients at the same
exponents in both sides of the last equality leads to the following
conditions on parameters $e_{s}$: 
\[
e_{s}^{*}=\prod_{S=1}^{2(l_{3}+l_{1})}(\lambda _{Ps}+\mu
_{S})\prod_{k=1,k\neq s}^{2l_{3}}(\lambda _{k}-\lambda _{s})^{-2}{\frac{1}{%
c_{(Ps)}}} 
\]

\subsection{Determinant computation. The case of $D$ function.}

Calculation of the determinant defining $D$ function and corresponding
condition of reality (\ref{AC}) does not add new limitations to the
numerical parameters obtained above in the last two subsections.

\section{Outlook}

No one of the authors is a specialist in the field of applications of 3-wave
problem to certain physical phenomenon. Due to this reason, we can not
discuss the usefulness of such applications.

From a computational point of view, it seems that here we have the first
paper having a systematical investigation of a multicomponent integrable
system. It is interesting to compare the results in the case of
multicomponent (the simplest one) integrable system connected with the
algebra $A_{2}$ and numerous integrable systems connected with $A_{1}$
algebra. The strategy of computations in both cases is basically the same.
Beginning from some simple solution connected with the upper triangular
nilpotent subalgebra ($(P=A=Q=0)$ in the case of the present paper after
corresponding $2n$ steps of discrete transformation, we come to solution
connected with the lower triangular nilpotent subalgebra. Connection with
hermitianity of these solutions leads to limitation of (arbitrary up to now)
numerical parameters of the problem. In the case of two component integrable
system, for instance in nonlinear Scrodinger equation, this dependence looks
as \cite{NLSC}. 
\[
c_{s}^{*}=\prod_{k=1,k\neq s}^{2l_{3}}(\lambda _{k}-\lambda _{s})^{-2}{\frac{%
1}{c_{(Ps)}}},\quad \lambda _{s}^{*}=-\lambda _{Ps} 
\]
In the case under consideration, this dependence modified by the formulas of
subsections $B,E$. This comparison offers possibility assuming that each
simple root of semisimple algebra borns its own systems of pairs of
parameters amplitude-phase and numbers of these pairs are arbitrary and only
by them multi-soliton solutions are defined in the case of multicomponent
integrable systems.

In our consideration, it was unable to connect constructed solutions with
L-A pair formalism and authors have no idea how this can be done (if
possible).

\section{Acknowledgements}

The authors thank CONACyT for financial support.


\begin{thebibliography}{99}
\bibitem{1}  Sukhorukov A.P. {\it Nonlinear interactions in optics and
radiophysics}, Nauka, Moskow (in Russian) (1988).

\bibitem{A}  WEILAND, J; WILHELMSSON, H {\it Coherent non-linear interaction
of waves in plasmas Oxford and New York, Pergamon Press (International
Series in Natural Philosophy. Vol. 88), pp. 366. 1977}

\bibitem{B}  Vlahos, L.; Sharma, R. R.; Papadopoulos, K. {\it Astrophysical
Journal, Part 1 (ISSN 0004-637X), vol. 275, Dec. 1, 1983, p. 374-390}

\bibitem{C}  O. MAHRENHOLTZ and M. MARKIEWICZ {\it Series: Advances in Fluid
Mechanics Vol 24 Published: 1999 Pages: 272pp}

\bibitem{D}  Kanemitsu Katou {\it Journal of the Physical Society of Japan
Vol. 51 No. 3, March, 1982 pp. 996-1000}

\bibitem{E}  K. Berger and U. Milevski {\it SIAM J. APPL. MATH. (2003)
Society for Industrial and Applied Mathematics, Vol, 63, No. 4, pp.1121-1140}

\bibitem{F}  Konotop VV, Cunha MD, Christiansen PL, Clausen CB.{\it Phys Rev
E Stat Phys Plasmas Fluids Relat Interdiscip Topics. 1999 Nov;60(5 Pt
B):6104-10}.

\bibitem{G}  L. TKESHELASHVILI,K. BUSCH {\it Appl. Phys. B 81, 225-229 (2005)%
}

\bibitem{I}  A.N.Leznov and R.Torres-Cordoba {\it J.Math.Phys (2002)}

\bibitem{2}  V.E.Zakharov, S.M.Manakov, S.P.Novikov and L.P.Pitaevskii {\it %
Theory of Solitons. The Method of the Inverse Scaterring Preoblem} Moscow,
Nauka, 1980 (in Russian)'

\bibitem{3}  A.N.Leznov {\it Theor. Math. Physica N123,v 2, (633-650), MAY
2000}.-

\bibitem{4}  A.N.Leznov {\it Physics of elementary particals and atom
nuclears} N27, v.5, p 1161-1246.

\bibitem{LS}  A.N.Leznov and M.V.Saveliev {\it Group methods for integration
of nonlinear dynamical systems} Progress in Physics 15, Basel, (1992)

\bibitem{LY}  A.N.Leznov and E.A.Yusbashjan {\it LMP v35, p. 345-349, (1995)}

\bibitem{LYY}  A.N.Leznov and E.A.Yusbashjan {\it Nucl.Phys.B
496,(3),643-653, (1997)}
\end{thebibliography}
\end{document}